\documentclass[12pt]{article}

\usepackage{amsmath}
\usepackage{amssymb}

\newtheorem{theorem}{Theorem}[section]
\newtheorem{lemma}{Lemma}[section]

\newtheorem{prop}{Proposition}[section]
\newtheorem{definition}{Definition}[section]
\numberwithin{equation}{section}

\begin{document}

\begin{center}

The C-version Segal-Bargmann transform \\
for finite Coxeter groups defined by the restriction principle

\end{center}

\vskip .4cm

\centerline{Stephen Bruce Sontz\footnote{Research partially
supported by CONACYT (Mexico) project 49187.}}

\centerline{Centro de Investigaci\'on en Matem\'aticas, A.C. (CIMAT)}

\centerline{Guanajuato, Mexico}

\centerline{email: sontz@cimat.mx}

\vskip .4cm

\begin{abstract}
\noindent
We apply a special case, the restriction principle (for which we give
a definition simpler than the usual one),
of a basic result in functional analysis
(the polar decomposition of an operator) in order to define $C_{\mu, t}$, the
$C$-version of the Segal-Bargmann transform,
associated to a finite Coxeter group acting in $\mathbb{R}^N$
and a given value $t>0$ of Planck's constant, where $\mu$~is a multiplicity function
on the roots defining the Coxeter group.
Then we immediately prove that $C_{\mu, t}$ is a unitary isomorphism.
To accomplish this we identify the reproducing kernel function of the appropriate Hilbert
space of holomorphic functions.
As consequences we prove that the Segal-Bargmann transforms for Versions $A$, $B$
and $D$ are also unitary isomorphisms, though not by a direct application of the
restriction principle.
The point is that the $C$-version is the the only version where a restriction
principle, in our definition of this method, applies directly.
This reinforces the idea that the $C$-version is the most fundamental, most
natural version of the Segal-Bargmann transform.
\end{abstract}



\section{Introduction}
\label{sec1}

The basic idea involved in the restriction principle is the use of the
polar decomposition of an operator in order to define a unitary transformation.
The polar  decomposition (e.g., see \cite{KA} and \cite{RS1}) is a well known result
in functional analysis that says that one can write $T = U |T| $, where $ |T| = ( T^* T)^{1/2} $
and $U$ is a partial isometry.
Here $T$ is a closed (possibly unbounded), densely defined linear operator
mapping its domain $\mathrm{Dom} (T) \subset \mathcal{H}_1 $ to $\mathcal{H}_2 $, where
$ \mathcal{H}_1$ and $ \mathcal{H}_2$ are complex Hilbert spaces.
It turns out that  $ ( T^* T)^{1/2} $ is the positive square root of the densely
defined self-adjoint operator  $  T^* T $ and so maps a domain in $ \mathcal{H}_1$
to $ \mathcal{H}_1$.
The partial isometry $U$ maps $ \mathcal{H}_1$
to $ \mathcal{H}_2$.
We are generally interested in the case when the partial isometry $U$
is a unitary isomorphism
from $ \mathcal{H}_1$ onto $ \mathcal{H}_2$, which is true if and only if $T$ is one-to-one
and has dense range.

Applying the polar decomposition theorem as a means for constructing unitary
operators is a very general method.
Also this method has nothing to do with the structures of the complex Hilbert spaces
$ \mathcal{H}_1$ and $ \mathcal{H}_2$.
And these can be advantages or disadvantages depending on one's particular interest.

But if we assume that $\mathcal{H}_2$ is some set of complex-valued functions
(and in general not equivalence classes of functions) 
on a set $X$  and $ \mathcal{H}_1$ is a Hilbert
space of complex-valued
functions (or possibly equivalence classes of functions)
on a subset $M$ of $X$,
then we define the \textit{restriction operator}
$ R : \mathcal{H}_2 \to \mathcal{H}_1$ by $Rf(x):= f(x)$
for all $f \in \mathcal{H}_2$ and all $x \in X$.
This is at a formal level only, since in general we do not
know that $Rf \, ( \, = f\upharpoonright_M = f$ restricted to $M)$ is an element
of $\mathcal{H}_1$.
Then we apply the polar decomposition to the adjoint $R^*$
of the restriction operator $R$
(\textit{provided} that $R^*$ is a closed, densely defined operator)
to get $ R^* = U P$ where $P$ is a positive operator of no further interest and
$U$ is a partial isometry from  $ \mathcal{H}_1$ to $ \mathcal{H}_2$.
We say that $U$ is defined by the \textit{restriction principle}.
We then have to show that $R^*$ is one-to-one and has dense range in order
to prove that the partial isometry $U$ is a unitary isomorphism
from $ \mathcal{H}_1$ onto $ \mathcal{H}_2$, this being the case of interest
for us.
In this paper $ \mathcal{H}_2$ will be a reproducing kernel Hilbert space.
This turns out to be quite useful for deriving explicit formulas, but is
not a necessary aspect of this approach.

We note that our definitions here differ from those of other authors.
For us a restriction operator is simply restriction to a subset and nothing else.
Other authors allow for operators that are the composition of restriction to a
subset followed or proceeded by another operator, often a multiplication operator.
Then these authors apply the polar decomposition
to these more general ``restriction'' operators.
Now this introduces another operator as a
\textit{deus ex machina}, that is, something that arrives on the stage
without rhyme or reason, but that saves the day by making everything work out well.
We object to such an approach to constructing a mathematical theory on general
principles, both aesthetic and logical.
Moreover, in the context of generalizations of Segal-Bargmann analysis it seems
that the application of the restriction principle (using our definition of
this) in the context of the $C$-version of Segal-Bargmann analysis eliminates
any need to introduce unmotivated factors.
This is clearly seen in this paper as well as in \cite{HZ} and \cite{OO}.
Also, as we shall see, developing the theory first for the $C$-version
gives us enough information to dispose easily of the other versions,
including an explanation of where the ``mysterious'' multiplication factors come from
for the $A$, $B$ and $D$ versions.
See Hall \cite{HA} for the original use of this nomenclature of ``versions''
and \cite{SBS2} for its use in the context of finite Coxeter groups.

When the above sketch can be filled in rigorously,
this is a simple way of
defining a unitary isomorphism $U$.
Moreover, the simplicity of the definition often allows one to prove results
about $U$ in a straightforward way.
However, the devil lies in the details as the saying goes, and the details
can sabotage this approach.
For example, the definition of the restriction operator $R$ might not make sense
on the domain $\mathcal{H}_2 $, though it always makes sense on the subspace
$\mathrm{Dom}(R):= \{ f \in \mathcal{H}_2 ~|~ Rf = f\upharpoonright_M
\in \mathcal{H}_1 \}$.
However, it could happen that $\mathrm{Dom}(R)$ is the zero subspace, in which
case this method is for naught.

The full history of this method is not our primary interest, but
we present what we know about this in the area of mathematical physics and
related areas of analysis.
In this paragraph, and only in this paragraph, the phrase ``restriction principle''
is used in the sense of the authors cited.
Peetre and Zhang in 1992 in \cite{PZ} used polarization
to get the Berezin transform.
A polar decomposition was used by \O{rsted} and Zhang in \cite{OZ} in order
to define and study the Weyl transform.
The article \cite{OO} by \'Olafsson and \O{rsted}
contains some applications of restriction principles
in order to understand the work of Hall in \cite{HA} and Hijab in \cite{HI}.
The approach in \cite{OO} was recently followed up by Hilgert and Zhang in \cite{HZ}
in their study of compact Lie groups.
Also Davidson, \'Olafsson and Zhang used a restriction principle in \cite{DOZ}
in order to study Laguerre polynomials.
See \cite{DOZ} for more references on this topic and on the Berezin transform.
Zhang in \cite{ZH} used a restriction principle to study the Segal-Bargmann transform
of a weighted Bergman space on a bounded symmetric domain.
In \cite{SBSBO} a restriction principle was used by Ben Sa{\"i}d and \O{rsted}
to produce a ``generalized Segal-Bargmann transform''
associated with a finite Coxeter group acting on $\mathbb{R}^N$.
We first learned about this method by reading \cite{OZ} within some
six months of its publication.
But our recent interest was stimulated by our desire to understand \cite{SBSBO}.

We should note that the same generalized Segal-Bargmann space as found in \cite{SBSBO}
together with its associated Segal-Bargmann transform (but called the
\textit{chaotic transform}) can be found for the case $M=\mathbb{R}$, $X=\mathbb{C}$
(dimension $N=1$) in Sifi and Soltani \cite{SISO}
and for $M=\mathbb{R}^N$, $X=\mathbb{C}^N$
(arbitrary finite dimension $N$) in Soltani \cite{SO}.
However, neither \cite{SISO} nor \cite{SO} used a restriction principle.
The case $M=\mathbb{R}$, $X=\mathbb{C}$
is discussed by us in \cite{SBS1} and in the references found there,
while we studied the arbitrary finite dimensional case $M=\mathbb{R}^N$, $X=\mathbb{C}^N$
in \cite{SBS2}.
Our point of view in \cite{SBS1} and \cite{SBS2} was to use
the approach of Hall \cite{HA}, which
is directly based on heat kernel analysis, rather than using the restriction principle.
While the restriction principle can be considered as an alternative to the approach
of Hall, this approach still relies in an essential way, at least in this paper,
on the heat kernel of the Dunkl theory as we shall see.

The restriction principle approach has various limitations.
For example, $X$ and $M$ need not be manifolds and, even if they are,
$X$ need not be the cotangent bundle of $M$ so that
the theory can lose contact with physics and symplectic geometry.
Also, the Hilbert spaces
are not constructed, but must be known prior to applying this approach.
And there is no necessary connection with heat kernel analysis.
Of course, these attributes can be viewed as strengths rather than weaknesses, since
they could allow for more general application than other approaches.

In this paper we will use the restriction principle to define the $C$~version of the
Segal-Bargmann transform $C_{\mu, t}$ associated
with a finite Coxeter group acting on $\mathbb{R}^N$
and with a value $t>0$ of Planck's constant.
(We will discuss the multiplicity function $\mu$ later on.)
We also show that $C_{\mu, t}$ is a unitary isomorphism.
This is a new way to construct $C_{\mu, t}$ and prove that it is a unitary isomorphism.
Along the way we have to find an explicit formula for the reproducing kernel function for
the Hilbert space $\mathcal{C}_{\mu, t}$ that turns out to be the range of the
unitary transform $C_{\mu, t}$.

A major point of this paper is that our original proof of the unitarity of
the transform $C_{\mu, t}$,
as given in \cite{SBS2}, depends on using the previously established unitarity
of $A_{\mu,t}$, the $A$-version of the Segal-Bargmann transform.
Since none of the versions of the Segal-Bargmann transform appears as the most
natural version in the analysis given in \cite{SBS2}, there is no logical reason
to start with the $A$-version.
However, using that approach, things in the end do work out quite nicely.
But the proof given here seems to us to be more natural, since the starting
point, namely the $C$-version, plays a distinguished role, while the remaining
versions are obtained as secondary constructs.

Having established these results in the $C$-version, it then is simple for us
to prove the corresponding results for Versions $A$, $B$ and $D$.
In particular we show as an immediate consequence to our work how the ``restriction''
operator used in \cite{SBSBO} (which is actually restriction followed
by multiplication by an unmotivated factor) arises in a natural way
from our restriction operator, which is simply restriction without
multiplication by some fudge factor.

The upshot is that the restriction principle for the $C$-version
can be used as a starting point for defining
all of the versions of the Segal-Bargmann transform associated to a finite Coxeter group.
Therefore the restriction principle is a fundamental principle in Segal-Bargmann analysis.
So, this paper complements the approach in our recent paper \cite{SBS2} where we showed
by using the Dunkl heat kernel that the versions~$A$,~$B$ and $C$
of the Segal-Bargmann transform associated with a finite Coxeter group are analogous
to the versions of the Segal-Bargmann transform as introduced by Hall in \cite{HA},
where he used the appropriate heat kernel.

Since many authors now take the $C$-version to be the most fundamental version
of the Segal-Bargmann transform, we feel that our result has an impact on that
approach to this field of research.
We also feel that the current approach is better than that in \cite{SBS2},
since we now emphasize how the $C$-version is singled out in yet another way
as more fundamental than the other versions.

\section{Definitions and other preliminaries}
\label{sec2}

We follow the definitions and notation of \cite{SBS2}.
Consult \cite{SBS2} and the references given there for a more
leisurely review of this material.
In that paper we studied various versions of the Segal-Bargmann transform associated with
a finite Coxeter group acting on the Euclidean space $\mathbb{R}^N$.
One of these versions (known as Version $A$ or the $A$-version) is,
as we shall see, a unitary isomorphism of Hilbert spaces,
$$
  A_{\mu,t} : L^2 ( \mathbb{R}^N, \omega_{\mu,t} ) \equiv L^2 ( \omega_{\mu,t} )
              \to \mathcal{B}_{\mu,t},
$$
where the density function (with respect to Lebesgue measure) 
for $q \in \mathbb{R}^n$ is
$$
 \omega_{\mu,t} (q) : = c_{\mu}^{-1} t^{-(\gamma_\mu + N/2)} \prod_{\alpha \in \mathcal{R}}
  | \left\langle \alpha , q \right\rangle |^{\mu(\alpha)}.
$$
Throughout this paper we let $t>0$ denote Planck's constant.
Here the {\em Macdonald-Mehta-Selberg constant} is defined by
$$
 c_\mu := \int_{\mathbb{R}^N} \mathrm{d}^{N} \! x \, \,
          t^{-(\gamma_\mu + N/2)} \, e^{-x^2/2t} \, \prod_{\alpha \in \mathcal{R}}
          | \left\langle \alpha , x \right\rangle |^{\mu(\alpha)}.
$$
(In a moment we shall discuss $\gamma_\mu$, the finite set $\mathcal{R}$ and
$\mu : \mathcal{R} \to [0, \infty)$.)
Since this integral does not depend on the value of $t>0$ (by dilating),
we do not include this parameter in the notation on the left side.
Clearly, $0 < c_\mu < \infty$.

We define the \textit{Version $A$ Segal-Bargmann transform} as the integral kernel operator
\begin{equation}
\label{def-sbt}
 A_{\mu,t} \psi (z) :=
\int_{\mathbb{R}^N} \mathrm{d} \omega_{\mu,t}(q) A_{\mu,t} (z,q) \psi(q)
\end{equation}
for $\psi \in L^2 ( \mathbb{R}^N , \omega_{\mu,t})$ and $z \in \mathbb{C}^N$ and $t>0$,
where the integral kernel is defined for $z \in \mathbb{C}^N$ and $q \in \mathbb{R}^N$ by
\begin{equation}
\label{def-kernel-a}
A_{\mu,t} (z,q) :=
\exp \left(-z^2/2t -q^2/4t\right)
E_{\mu} \left( \dfrac{z}{ t^{1/2} },\dfrac{q}{ t^{1/2} }  \right),
\end{equation}
where
$z^2 := z_1^2 + \cdots + z_N^2 $ (given that $z=(z_1, \dots, z_N))$
is a holomorphic function and $q^2 := ||q||^2$ is the usual Euclidean norm squared.
The function $E_\mu : \mathbb{C}^N \times \mathbb{C}^N \to \mathbb{C}$
will be introduced momentarily.
Using the inequality~(\ref{E_mu_estimate}) below and the Cauchy-Schwarz inequality
one shows the absolute convergence of the integral in (\ref{def-sbt}).
Our paper \cite{SBS2} provides motivation for formula (\ref{def-kernel-a}).

In the above $\mathcal{R}$ is a certain finite subset of  $\mathbb{R}^N$,
known as a \textit{root system},
$\mu : \mathcal{R} \to [0,\infty) $ is a \textit{multiplicity function}
(see \cite{SBS2} for definitions)
and
$$
    \gamma_{\mu} := \dfrac{1}{2} \sum_{\alpha \in \mathcal{R}} \mu (\alpha).
$$
It may be possible to weaken the hypothesis $\mu \ge 0$ that we are imposing here while
still having the same results.
We work with a fixed root system  $\mathcal{R}$ and a fixed multiplicity function $\mu$
throughout this article.
See \cite{SBS2} for the details about how $\mathcal{R}$ gives rise to a finite
Coxeter group acting as orthogonal transformations of $\mathbb{R}^N$.

The space $ \mathcal{B}_{\mu,t}$ introduced above (\cite{SBSBO}, \cite{SO}),
which is called the \textit{Version~$A$ Segal-Bargmann
space}, is the reproducing kernel Hilbert space of holomorphic functions
$f : \mathbb{C}^N \to \mathbb{C}$ whose \textit{reproducing kernel}
$
      K_{\mu,t} :  \mathbb{C}^N \times  \mathbb{C}^N \to  \mathbb{C}
$
is defined for $z,w \in \mathbb{C}^N $ and $t>0$ by
$$
K_{\mu,t}(z,w) = E_\mu \left( \frac{z^*}{t^{1/2}}  , \frac{w}{t^{1/2}}  \right),
$$
where $E_\mu$ is the \textit{Dunkl kernel function} associated with the Coxeter group
 (associated itself to the root system $\mathcal{R}$) and the multiplicity function $\mu$.
For any $z=(z_1, \dots , z_N) \in \mathbb{C}^N $ we let
$z^*=(z_1^*, \dots , z_N^*) \in \mathbb{C}^N$
denote its complex conjugate.
The Dunkl kernel $E_\mu : \mathbb{C}^N \times \mathbb{C}^N \to \mathbb{C}$
(\cite{dJ}, \cite{DU}, \cite{RO2})
is a holomorphic function with many properties.
We simply note for now that
\begin{eqnarray*}
       E_\mu (z, 0) &=& 1
\\
     E_\mu (z, w) &=& E_\mu (w,z)
\\
     E_\mu (\lambda z, w) &=& E_\mu (z, \lambda w)
\\
     (E_\mu (z, w))^* &=& E_\mu (z^* , w^*) \\
  E_\mu (z, w) &=& \exp (z \cdot w) = e^{ z \cdot w } \quad \mathrm{if} \quad  \mu \equiv 0
\end{eqnarray*}
for all $\lambda \in \mathbb{C}$ and all $z,w \in \mathbb{C}^N$.
In the first equation $0$ denotes the zero vector in $\mathbb{C}^N$.
Also, $z \cdot w = \sum_j z_j w_j $ in the obvious notation.
We will also be using the estimate (see \cite{RO1})
\begin{equation}
\label{E_mu_estimate}
      | E_\mu( z, w) | \le \exp ( ||z|| \, ||w|| )
\end{equation}
for all $z,w \in \mathbb{C}^N$, which holds if $\mu \ge 0$.
(Here, $|| z ||$ is the Euclidean norm of $z \in \mathbb{C}^N$.
Also recall that $\mu \ge 0$ is assumed
throughout this article.)

For a Hilbert space $\mathcal{H}$ we use the notations
$\langle \cdot , \cdot \rangle_{ \mathcal{H} }$ and
$|| \cdot ||_{ \mathcal{H} }$ for its inner product and norm, respectively.
The inner product is anti-linear in its first argument, linear in its second.
All Hilbert spaces considered are over the field of complex numbers.

We will be using dilations.
Our present notation for these operators is
$D_\lambda \psi (x) := \psi (\lambda x)$, where
$\psi$ is a function in some appropriate function space.
The proof of the next result is straightforward and so is left to the reader.
\begin{lemma}
\label{dilation_lemma}
 For every $\lambda >0 $ and $t>0$, we have that
$$
  \lambda^{\gamma_{\mu} +N/2   }  D_\lambda :  L^2 ( \mathbb{R}^N, \omega_{\mu,t} ) \to
                        L^2 ( \mathbb{R}^N, \omega_{\mu,t} )
$$
is a unitary isomorphism.
\end{lemma}

Finally, we want to introduce the Dunkl heat kernel
(see \cite{RO1} and \cite{RO2}) for the heat equation
associated with the \textit{Dunkl Laplacian} $\Delta_{\mu} $, namely
\begin{equation}
\label{dunkl_heat_eqn}
        \dfrac{\partial u}{\partial t} = \dfrac{1}{2} \Delta_{\mu} u.
\end{equation}
The Dunkl Laplacian $\Delta_{\mu} $ is defined and discussed in \cite{RO1}.
In particular, it has a realization in $L^2 ( \mathbb{R}^N, \omega_{\mu,t} )$
as an unbounded, self-adjoint  operator with $\Delta_{\mu} \le 0$ 
and spectrum $(- \infty, 0]$.
Specifically, we have for $t>0$ and $x \in \mathbb{R}^N$ that
$$
 u(x,t) = e^{t \Delta_\mu/2} f (x) =
          \int_{ \mathbb{R}^N } \mathrm{d} \omega_{\mu, t} (q) \rho_{\mu, t} (x, q) f(q)
$$
solves (\ref{dunkl_heat_eqn}) for any initial condition
$f \in L^2 ( \mathbb{R}^N, \omega_{\mu,t} )$
(see \cite{RO2} for more details),
where the \textit{Dunkl heat kernel}
$\rho_{\mu,t}: \mathbb{R}^N \times \mathbb{R}^N \to \mathbb{R}$ is given
 for all $x,q \in \mathbb{R}^N$ and $t>0$ by
\begin{equation}
\label{define_rho}
    \rho_{\mu,t} (x,q) = e^{-(x^2 + q^2)/2t}
            E_\mu \left( \dfrac{x}{t^{1/2}}, \dfrac{q}{t^{1/2}}  \right).
\end{equation}
This has an analytic extension  $ \mathbb{C}^N \times \mathbb{C}^N \to \mathbb{C}$,
which we also denote as $\rho_{\mu,t}$.
One of the basic results of \cite{SBS2} is that for $z \in \mathbb{C}^N$ and
$q \in \mathbb{R}^N$ we have
$$
   A_{\mu, t} (z,q) = \dfrac{ \rho_{\mu, t} (z,q) }{ ( \rho_{\mu, t} (0,q) )^{1/2} }
$$
which, in accordance with the approach of Hall \cite{HA}, indicates that
(\ref{def-sbt}) is justifiably called the Version~$A$ Segal-Bargmann transform
associated with a finite Coxeter group.
This formula also clarifies the nature of the seemingly arbitrary definition
 (\ref{def-kernel-a}) of the kernel function of the integral transform $A_{\mu, t}$.

Notice that the reproducing kernel function for $\mathcal{B}_{\mu,t}$ clearly satisfies
$$
   K_{\mu, t} (z,w) = E_\mu \left( \dfrac{z^*}{t^{1/2}}, \dfrac{w}{t^{1/2}} \right) =
\dfrac{ \rho_{\mu, t} (z^*,w) }{ \rho_{\mu, t} (z^*,0) \rho_{\mu, t} (0,w)   }.
$$
This identity shows that the reproducing kernel function for
the Hilbert space $\mathcal{B}_{\mu,t}$
is determined by the Dunkl heat kernel $\rho_{\mu, t}$.
Or, in other words, we can get the Segal-Bargmann space for Version~$A$
from the Dunkl heat kernel.
Another way to write this reproducing kernel in terms of the Dunkl heat kernel
$\rho_{\mu, t}$ is to consider equation (46) in Hall \cite{HA}.
In the present context the analogous result says that for
all $z, w \in \mathbb{C}^N$ we have
$$
   K_{\mu, t} (z,w) = \int_{ \mathbb{R}^N } \! \! \mathrm{d} \omega_{\mu, t} (q) \,
   \dfrac{\rho_{\mu, t}(w,q)  \rho_{\mu, t}(z,q)^* }{ \rho_{\mu, t}(q,0) }
$$
as the reader can check.
(Hint: One needs an identity involving the Dunkl kernel. See \cite{SBSBO}, equation (2.5),
or \cite{RO2}, Proposition 2.37, equation (2).)
Even though we will not be using these two formulas for $K_{\mu, t} (z,w)$, we present
them to show how the Dunkl heat kernel determines the reproducing kernel of
$\mathcal{B}_{\mu,t}$.
As we shall show later, the reproducing kernel function for the Version~$C$ Segal-Bargmann
space is also determined by the Dunkl heat kernel $\rho_{\mu, 2t}$.

We gather here some basic results of functional analysis that we will be using.
(See \cite{KA}, especially Chap.~III, \S 5 and Chap.~V, \S 3, for more details.)
 Let $\mathcal{H}_1$ and $\mathcal{H}_2$ be complex Hilbert spaces with
$T : \mathrm{Dom} (T) \to \mathcal{H}_2$ a linear operator which is densely defined
(which means $\mathrm{Dom} (T)$ is a dense subspace in $\mathcal{H}_1$).
Let $T^*$ denote the adjoint of $T$.
If $T$ is closable (namely, has a closure), then we denote the closure
of $T$ by $\overline{T}$.
We denote the kernel and range of $T$ by $ \mathrm{Ker} \, T$ and $ \mathrm{Ran} \, T $,
respectively.
We say that $T$ is \textit{globally defined} if $\mathrm{Dom} (T) = \mathcal{H}_1$.
For any subset $A$ in a Hilbert space, $\overline{A}$ is its closure in the norm topology
and $A^\perp$ is its orthogonal complement.
The following proposition comes from elementary functional analysis.

\begin{prop}
\label{fa_prop}

Let $T : \mathrm{Dom} (T) \to \mathcal{H}_2$ be densely defined, as above.
Then we have the following.
\begin{enumerate}
 \item If $T$ is closable, then $T^*$ is closed, densely defined and $\overline{T}= T^{**}$.
\item $\mathrm{Ker} \, T^* = ( \mathrm{Ran} \, T )^\perp$.
\item If $T$ is closed,
      then $\overline{ \mathrm{Ran} \, T^* } = ( \mathrm{Ker} \, T )^\perp$.
\item If $T$ is bounded (i.e., there exists $C \ge 0$ such that
$|| T \phi ||_{ \mathcal{H}_2 } \le C || \phi ||_{ \mathcal{H}_1 }$
for all $\phi \in \mathrm{Dom} (T) \,$), then $T$ is closable and
$\overline{T}$ is globally defined and bounded (with the same bound as $T$).
In particular, if $T$ is bounded and closed, then $T$
is globally defined, that is,  $ \mathrm{Dom} (T) = \mathcal{H}_1 $.
\end{enumerate}
\end{prop}

As we have already mentioned, we will use a standard result of functional analysis
known as the polar decomposition of an operator.
For the reader's convenience we state this result.
We present a modification of the statement of Theorem VIII.32 in \cite{RS1}.
A very thorough discussion of this topic is also given in \cite{KA}.
(See Chap.~VI, \S 2.7.)
We state this theorem for a closed densely defined linear operator
(that is, it may be bounded or not).

\begin{theorem}
\label{polar_decomp}
 (Polar Decomposition)
Let $\mathcal{H}_1$ and $\mathcal{H}_2$ be Hilbert spaces and
$A : \mathrm{Dom}(A) \to \mathcal{H}_2 $ be a closed linear operator, defined
in the dense linear domain $\mathrm{Dom}(A) \subset \mathcal{H}_1$.
Then there exists a positive self-adjoint operator $|A| := (A^* A)^{1/2}$
with $\mathrm{Dom}(|A|) = \mathrm{Dom}(A)$
and there exists a partial isometry $U : \mathcal{H}_1 \to \mathcal{H}_2 $ with initial
space $(\mathrm{Ker} \, U )^\perp = (\mathrm{Ker} \, A )^\perp $
and final space $\mathrm{Ran} \, U = \overline{ \mathrm{Ran} \, A } $ such that
$$
      A = U \, |A|
$$
on their common domain $\mathrm{Dom}(A) = \mathrm{Dom}(|A|)$.
Also, $U$ and $|A|$ are uniquely determined by $ \mathrm{Ker} \, |A| = \mathrm{Ker} \, A $
and the above properties.

In particular, $U$ is one-to-one if and only if $\mathrm{Ker} \, A = 0$, while
$U$ is onto if and only if $\mathrm{Ran} \, A$ is dense.

Consequently, $U$ is a unitary isomorphism of $\mathcal{H}_1$ onto $\mathcal{H}_2$ 
if and only if $\mathrm{Ker} \, A = 0$ and $\mathrm{Ran} \, A$ is dense.
\end{theorem}

\noindent
\textbf{Remarks:} 
Theorem~\ref{polar_decomp} is stated in terms of the structures of Hilbert spaces,
nothing else.
So it is invariant under unitary isomorphisms.
To make this more explicit we suppose
$F_j :\mathcal{H}_j \to \mathcal{K}_j$ are unitary
isomorphisms for $j=1,2$, where $\mathcal{K}_1$ and $\mathcal{K}_2$ are Hilbert spaces.
(We continue using the notation of Theorem \ref{polar_decomp}.)
Then define $\mathrm{Dom} (B):= F_1(\mathrm{Dom} (A))$, a subset of $\mathcal{K}_1$,
and $ B : \mathrm{Dom} (B) \to \mathcal{K}_2$ by $B:= F_2 A F_1^*$.
Clearly, $B$ is a closed, densely defined operator.
So, according to Theorem~\ref{polar_decomp}, we have that $B = V |B| $,
where $|B| = (B^* B)^{1/2}$ and $V : \mathcal{K}_1 \to \mathcal{K}_2$
is a uniquely determined partial isometry.
Then the relation of the polar decomposition of $B$ with that of $A = U |A|$ is
\begin{equation}
 \label{unitarty_trans}
|B| = F_1 |A| F_1^* \quad  \mathrm{and}  \quad V = F_2 U F_1^*.
\end{equation}
Moreover, if $A = R^*$ where $R$ is a restriction operator, then
according to our definition
$U$ is defined by a restriction principle.
Nonetheless,
$V$ need not be defined by a restriction principle, that is, $B$ need not
be the adjoint of a restriction operator even though $A$ is.
However, $V$ is well defined by polar
decomposition.
While the restriction principle is not a unitary invariant, this discussion
shows that there is a straightforward method for transforming a polar decomposition
by unitary transformations.
There is absolutely no guesswork involved.

It seems to be a rule of thumb in Segal-Bargmann analysis
that it is rather straightforward to prove that a Segal-Bargmann transform
is injective, while to prove that it is surjective requires a rather detailed argument.
However, that is not so for the restriction principle we will consider.
On the contrary, as we shall see in the next section, proving that the transform is
surjective is immediate (using uniqueness of analytic continuation), while
proving that it is injective does involve a bit more work
(using that the Dunkl transform, to be discussed later,
is injective), though is not all that difficult.

\section{Version $C$}

In this section we shall show how Version $C$ of
the Segal-Bargmann transform associated to a Coxeter group
arises from the restriction principle.
We feel that using the restriction principle is a more fundamental approach to this theory.

We recall from \cite{SBS2} that the \textit{Version $C$ (or $C$-version)
Segal-Bargmann transform}
for $\psi \in L^2( \mathbb{R}^N , \omega_{\mu,t})$ and $z \in \mathbb{C}^N$ is defined by
\begin{equation}
\label{def_C}
C_{\mu,t} \psi (z) := \int_{ \mathbb{R}^N }
       \mathrm{d} \omega_{\mu,t}(q) C_{\mu,t}(z,q) \psi (q),
\end{equation}
where $C_{\mu,t}(z,q) := \rho_{\mu, t}(z,q)$.
This integral converges absolutely by using the estimate (\ref{E_mu_estimate}).
This definition is the natural analogue in this context of the definition
of the $C$-version given in \cite{HA}.
Then we proved in \cite{SBS2} that this gives a unitary isomorphism
$$
C_{\mu, t} : L^2( \mathbb{R}^N , \omega_{\mu,t}) \to \mathcal{C}_{\mu,t}.
$$
The definition of the Hilbert space $\mathcal{C}_{\mu,t}$ of
holomorphic functions will be given below.
Other details may be found in \cite{SBS2}.
First, we will identify the reproducing kernel function for this Hilbert space.
Actually, this has already been done in \cite{SBS3} but the proof there
used the unitarity of the Version $C$ Segal-Bargmann transform $C_{\mu, t}$.
Since we wish here to construct that transform and then show that it is unitary,
we present an independent proof of this result.
So we consider the formula in \cite{SBS3} only as motivation for the
definition in equation (\ref{defineL}) of the following theorem,
which we now procede to prove without any reference to the transform $C_{\mu, t}$.

We again call to the reader's attention that restriction principles do not define
the Hilbert spaces, which must be introduced prior to the application of a restriction
principle.
And so it is in the present case with the Hilbert space $\mathcal{C}_{\mu,t}$.

\begin{theorem}
\label{thm41}
The reproducing kernel function $L_{\mu,t}$ for the Hilbert space $\mathcal{C}_{\mu,t}$ is given by
\begin{equation}
\label{defineL}
L_z (w)=
L(z,w) = L_{\mu,t}(z,w) :=  2^{-(\gamma_\mu + N/2)} \rho_{\mu, 2t} (z^*, w)
\end{equation}
for all $ z, w \in \mathbb{C}^N$.
\end{theorem}
\textbf{Remarks:}
Note the similarity of formula (\ref{defineL}) with the reproducing kernel
for the Version $C$ generalized Segal-Bargmann space for compact, connected
Lie groups as given by Hall in \cite{HA} (Theorem~6, p.~127):
$$
        \rho_{2t} (g^{-1} \overline{h}) \quad \quad  g,h \in G.
$$
See \cite{HA} for the definition of this notation and further details.
Also, note that this formula occurs in Segal-Bargmann analysis
in the context of Heisenberg groups
in \cite{RR} and in the context of the compact Heckman-Opdam setting in \cite{KTY}.
Admittedly, the factors of $2$ in our formula look strange and are not
found in these references.
These factors are a consequence of the unusual convention we have introduced in \cite{SBS2}
for normalizing the Dunkl heat kernel $\rho_{\mu, t}$
and the measure $\mathrm{d} \omega_{\mu, t}$.

\vskip 0.2cm
\noindent
\textbf{Proof:}
We let $\mathcal{H} (\mathbb{C}^N)$ denote the space of all of the
holomorphic functions $f: \mathbb{C}^N \to \mathbb{C}$.
We recall three definitions from \cite{SBS2}.
For $f \in \mathcal{H} (\mathbb{C}^N)$ we define $Gf \in \mathcal{H} (\mathbb{C}^N)$
by
\begin{equation}
\label{define_G}
   Gf(w) := 2^{\gamma_\mu/2 + N/4} f(2w) \, / \, A_{\mu,2t}(2w,0)
\end{equation}
for all $w \in \mathbb{C}^N$.
(Note that $A_{\mu,2t}(2w,0) = \exp (-w^2/t)$ is never zero.)
Then we define
\begin{equation}
\label{define_calC}
 \mathcal{C}_{\mu,t}:= \{ f \in \mathcal{H} (\mathbb{C}^N) ~|~ Gf \in \mathcal{B}_{\mu,t/2} \},
\end{equation}
which becomes a Hilbert space with its inner product defined by
\begin{equation}
\label{define_IP}
  \langle f_1, f_2 \rangle_{ \mathcal{C}_{\mu,t}} :=
   \langle Gf_1 , Gf_2 \rangle_{\mathcal{B}_{\mu,t/2}}
\end{equation}
for $f_1, f_2 \in \mathcal{C}_{\mu,t}$.

The reproducing kernel of a Hilbert space must satisfy two characteristic properties.
The first of these is that
$
     L_z ( \cdot) = 2^{-(\gamma_\mu + N/2)} \rho_{\mu, 2t} (z^*, \, \cdot)
$
must be an element in the Hilbert space $\mathcal{C}_{\mu,t}$.
The second is that
$
     f(z) = \left\langle L_z, f \right\rangle_{ \mathcal{C}_{\mu,t} }
$
for all $f \in \mathcal{C}_{\mu,t}$ and $z \in \mathbb{C}^N$.

We start with the first property.
Now $ L_z \in \mathcal{C}_{\mu,t}$ if and only if
$$
GL_z (w) =  2^{\gamma_\mu/2 + N/4} L_z(2w) \, / \, A_{\mu,2t}(2w,0)
$$
is an element of $\mathcal{B}_{\mu,t/2}$ as a function of $w \in \mathbb{C}^N$.

\noindent
So we calculate:
\begin{eqnarray*}
&&GL_z (w)
= 2^{\gamma_\mu/2 + N/4} L_z(2w) \, / \, A_{\mu,2t}(2w,0)
\\
 &=&
 2^{ -(\gamma_\mu/2 + N/4) } \rho_{\mu, 2t} (z^*, 2w) \, \!\exp (w^2/t)
\\
&=& 2^{ -(\gamma_\mu/2 + N/4) } \exp \left( \dfrac{-(z^*)^2 - 4w^2}{4t}  \right)
    E_\mu \left( \dfrac{z^*}{(2t)^{1/2}}  , \dfrac{2w}{(2t)^{1/2}}   \right)  \exp (w^2/t)
\\
&=& 2^{ -(\gamma_\mu/2 + N/4) } \exp( -(z^*)^2 /4t ) \,
                   E_\mu \left( \dfrac{z^*}{(2t)^{1/2}}  , \dfrac{2w}{(2t)^{1/2}}   \right)
\\
&=& 2^{ -(\gamma_\mu/2 + N/4) } \exp( -(z^*)^2 /4t ) \,
         E_\mu \left( \dfrac{z^*/2}{(t/2)^{1/2}}  , \dfrac{w}{(t/2)^{1/2}}   \right)
\\
&=& 2^{ -(\gamma_\mu/2 + N/4) } \exp( -(z^*)^2 /4t ) \,
                   K_{\mu, t/2} (z/2, w).
\end{eqnarray*}
Here $ K_{\mu, t/2} $ is the reproducing kernel function for
the Hilbert space $\mathcal{B}_{\mu,t/2}$, which
implies that $K_{\mu, t/2} (z/2, \, \cdot ) \in \mathcal{B}_{\mu,t/2}$
for all $z \in \mathbb{C}^N$
and so $GL_z \in  \mathcal{B}_{\mu,t/2} $ as desired.

Now for the second property
$
     f(z) = \left\langle L_z, f \right\rangle_{ \mathcal{C}_{\mu,t} }
$
we evaluate the right side for $f \in  \mathcal{C}_{\mu,t} $
(which implies $Gf \in \mathcal{B}_{\mu,t/2} $) and use $GL_z \in  \mathcal{B}_{\mu,t/2}$ to get
\begin{eqnarray*}
  \left\langle L_z, f \right\rangle_{ \mathcal{C}_{\mu,t} }
&=& \left\langle GL_z , Gf \right\rangle_{\mathcal{B}_{\mu,t/2}}
\\
&=&  \left\langle 2^{ -(\gamma_\mu/2 + N/4) } \exp( -(z^*)^2 /4t ) \,
                   K_{\mu, t/2} (z/2, \, \cdot) ,
 Gf \right\rangle_{\mathcal{B}_{\mu,t/2}}
\\
&=& 2^{ -(\gamma_\mu/2 + N/4) } \exp (-z^2/4t) \,
               \left\langle K_{\mu, t/2} (z/2, \,  \cdot ) ,
 Gf \right\rangle_{\mathcal{B}_{\mu,t/2}}
\\
&=& 2^{ -(\gamma_\mu/2 + N/4) } \exp (-z^2/4t) \,
Gf (z/2 )
\\
&=& 2^{ -(\gamma_\mu/2 + N/4) } \exp (-z^2/4t) \,
2^{\gamma_\mu/2 + N/4} f(z) \, / \, A_{\mu,2t}(z,0)
\\
&=& \exp (-z^2/4t) \, f(z) \, / \, A_{\mu,2t}(2(z/2),0)
\\
&=& \exp (-z^2/4t) \, f(z) \, \exp (z^2/4t) 
\\
&=& f(z)
\end{eqnarray*}
for all $z \in \mathbb{C}^N$.
So the second property has also been established, thereby completing the proof
without ever using the transform $C_{\mu, t}$.
$\blacksquare$

The definition of the Hilbert space $\mathcal{C}_{\mu,t}$ given in
(\ref{define_G}), (\ref{define_calC}) and (\ref{define_IP}) is what we were naturally led to while preparing \cite{SBS2}.
It is the range space of the Version~$C$ Segal-Bargmann transform $C_{\mu, t}$
introduced there.
However, the result of Theorem \ref{thm41} gives us an intrinsic way of defining
$\mathcal{C}_{\mu,t}$, namely as the Hilbert space of holomorphic functions
$f : \mathbb{C}^N \to \mathbb{C}$ with reproducing kernel defined by (\ref{defineL}).
This is arguably a better approach.
However, the natural way to do this would be to omit the factors of $2$ from
(\ref{defineL}).
This would simply give us a different normalization of the Version~$C$
of the Segal-Bargmann space.
But either way the Hilbert space $\mathcal{C}_{\mu,t}$ must be defined before
applying a restriction principle, as we noted earlier.

Of course, in order to apply the restriction principle,
we need to define the restriction operator rigorously.

\begin{definition}
We define the {\em restriction operator}
$$
R \equiv R_{\mu, t} : \mathrm{Dom}(R_{\mu, t}) \to  L^2(\mathbb{R}^N , \omega_{\mu,t})
$$
by
$$
(R_{\mu, t}f)(x) := f(x)$$
for all $f$ in a domain $\mathrm{Dom}(R_{\mu, t}) \subset \mathcal{C}_{\mu,t} $
and all $x \in \mathbb{R}^N$.
The definition of the domain of $R_{\mu,t}$ in $\mathcal{C}_{\mu,t}$ is the obvious one:
$$
\mathrm{Dom}(R) = \mathrm{Dom}(R_{\mu, t}) :=
\{ f \in  \mathcal{C}_{\mu,t} ~|~ f \upharpoonright_{\mathbb{R}^N} \in L^2( \mathbb{R}^N, \omega_{\mu,t} ) \}.
$$
Note that $R_{\mu, t}$ does depend on $\mu$ and $t$, since these parameters appear
in both the domain and codomain spaces of this operator.
\end{definition}

We will show later on that $R_{\mu,t}$ is a globally defined, bounded operator.
Still this is a bit surprising since the following standard estimates
do \textit{not} prove it.
Indeed, for any $0 \ne f \in \mathrm{Dom}(R_{\mu, t})  \subset \mathcal{C}_{\mu,t}$ we have that
\begin{eqnarray*}
&&|| R_{\mu, t} f ||^2_{L^2 (\omega_{\mu, t} ) } =
\int_{ \mathbb{R}^N }  \mathrm{d} \omega_{\mu, t}(x) \, | R_{\mu, t} f(x)|^2
= \int_{ \mathbb{R}^N } \mathrm{d} \omega_{\mu, t} (x) \, | f(x)|^2
\\
&\le& \int_{ \mathbb{R}^N } \! \mathrm{d}  \omega_{\mu, t} (x) \,
L_{\mu, t} (x,x) ||f||^2_{ \mathcal{C}_{\mu, t} }
=  \int_{ \mathbb{R}^N } \! \mathrm{d} \omega_{\mu, t} (x) \,
2^{-(\gamma_\mu + N/2)} \rho_{\mu, 2t} (x, x)
       ||f||^2_{ \mathcal{C}_{\mu, t} }
\\
&=& 2^{-(\gamma_\mu + N/2)} \int_{ \mathbb{R}^N } \mathrm{d} \omega_{\mu, t} (x) \,
e^{-(x^2 + x^2)/4t}
    E_{\mu} \left( \dfrac{x}{(2t)^{1/2}} , \dfrac{x}{(2t)^{1/2}} \right)
||f||^2_{ \mathcal{C}_{\mu, t} }
\\
&\le& 2^{-(\gamma_\mu + N/2)} \int_{ \mathbb{R}^N } \mathrm{d} \omega_{\mu, t} (x) \,
 e^{-x^2/2t} e^{x^2/2t} 
       ||f||^2_{ \mathcal{C}_{\mu, t} }
\\
&=& 2^{-(\gamma_\mu + N/2)}  \int_{ \mathbb{R}^N } \mathrm{d} \omega_{\mu, t} (x) \,
||f||^2_{ \mathcal{C}_{\mu, t} }
 = + \infty.
\end{eqnarray*}
Here we used (\ref{E_mu_estimate}) in the second inequality, and
the usual pointwise estimate for functions in a reproducing kernel Hilbert space
in the first inequality.

As far as we know at this point of our exposition it could well be the case
that $\mathrm{Dom}(R_{\mu, t}) = 0$.
We now show that this domain is actually dense along with other properties of $R_{\mu, t}$.

\begin{theorem}
\label{thm42}
The operator $R \equiv R_{\mu,t} $ defined on its domain
$\mathrm{Dom}(R_{\mu, t})$ is a closed, densely
defined operator that is one-to-one and has dense range in
$L^2( \mathbb{R}^N, \omega_{\mu,t} )$.
Also its adjoint $R_{\mu,t}^* $ is densely defined, closed, one-to-one and has dense range.
In particular, we have that $L_z \in \mathrm{Dom}(R_{\mu, t}) $
for all  $z \in \mathbb{C}^N$.
\end{theorem}
\textbf{Proof:}
By the uniqueness of analytic continuation from $\mathbb{R}^N$ to $\mathbb{C}^N$, we have
immediately that $R_{\mu,t}$ is one-to-one, that is, $\mathrm{Ker} \, R_{\mu, t} = 0$.

We claim that the functions $L_z \in \mathcal{C}_{\mu, t} $ are all
in  $\mathrm{Dom}(R)$. This follows from the equalities
\begin{eqnarray*}
   L_z(x) &=& 2^{-(\gamma_\mu + N/2)} \rho_{\mu, 2t} (z^*, x )
\\
 &=& 2^{-(\gamma_\mu + N/2)} \exp \left( \dfrac{-(z^*)^2 - x^2}{4t} \right)
            E_\mu \left( \dfrac{z^*}{(2t)^{1/2}} , \dfrac{x}{(2t)^{1/2}} \right)
\end{eqnarray*}
for $z \in \mathbb{C}^N$ and $x \in \mathbb{R}^N$,
which (using $\mu \ge 0$ and (\ref{E_mu_estimate})) give the estimate
$$
| L_z(x) | \le 2^{-(\gamma_\mu + N/2)} \exp \left( \dfrac{-Re (z^*)^2}{4t} \right)
\exp \left( \dfrac{- x^2}{4t} \right) \exp \left( \dfrac{||z^*|| \, ||x||}{2t} \right).
$$
This clearly implies that $| L_z(x) |^2  $ is integrable
with respect to the measure $\mathrm{d} \omega_{\mu,t}(x)$.
And so $L_z \in \mathrm{Dom}(R) $.
Now, by the theory of reproducing kernel Hilbert spaces,
the finite linear combinations of the functions $L_z$ with $z \in \mathbb{C}^N$
form a dense subspace of $\mathcal{C}_{\mu, t} $ and so  $\mathrm{Dom}(R)$ is dense, that is,
$R$ is a densely defined operator.

The proof that the graph of $R$ is closed is a standard argument,
which we leave to the reader.
So, $R$ is a closed operator.

The proof that $R^*$ is a densely defined and closed operator follows
by applying Proposition \ref{fa_prop} to the closed operator $R$.

To prove that $R_{\mu, t}^*$ is injective, we
first find a formula for $R_{\mu, t}^*$.
So we take
$\psi \in \mathrm{Dom} (R^*) \subset L^2(\omega_{\mu,t})$ and $z \in \mathbb{C}^N$
with the intention of calculating $R_{\mu, t}^* \psi (z)$ in general.
Introducing the reproducing kernel $L_z$ in the second equality
and using $L_z \in \mathrm{Dom}(R) $
in the third equality we calculate as follows:
\begin{eqnarray}
&&R_{\mu, t}^* \psi (z) =
 R^* \psi (z) = \left\langle L_z , R^* \psi \right\rangle_{ \mathcal{C}_{\mu,t} }
= \left\langle R L_z ,  \psi \right\rangle_{ L^2 ( \omega_{\mu,t} ) }
\nonumber
\\
&=& \int_{ \mathbb{R}^N } \mathrm{d} \omega_{\mu, t}(q) \, ( R L_z (q) )^* \psi (q)
= \int_{ \mathbb{R}^N } \mathrm{d} \omega_{\mu, t}(q) \, ( L_z (q) )^* \psi (q)
\nonumber
\\
&=&  \int_{ \mathbb{R}^N } \! \mathrm{d} \omega_{\mu, t}(q) \, 2^{-(\gamma_\mu + N/2)}
 \left( \rho_{\mu, 2t } (z^*, q ) \right)^* \psi (q)
= \int_{ \mathbb{R}^N } \! \mathrm{d} \omega_{\mu, 2t}(q) \, \rho_{\mu, 2t } (z, q ) \psi (q)
\nonumber
\\
&=& \int_{ \mathbb{R}^N } \! \! \mathrm{d} \omega_{\mu, 2t}(q) \,
e^{-z^2/4t} e^{-q^2/4t} E_\mu \left( \dfrac{z}{(2t)^{1/2}}, \dfrac{q}{(2t)^{1/2}}  \right)
 \psi (q).
\label{Rstar}
\end{eqnarray}
Notice how the factors of $2$ combined with $ \mathrm{d} \omega_{\mu,t}$
to form $ \mathrm{d} \omega_{\mu,2t}$,
which is the measure we want to use in integrals involving the
Dunkl heat kernel $\rho_{\mu, 2t}$.
Now put $z = -ix$ for $x \in \mathbb{R}^N$ in (\ref{Rstar}) to get
\begin{gather}
 R_{\mu, t}^* \psi (-ix) =  \int_{ \mathbb{R}^N }  \mathrm{d} \omega_{\mu, 2t}(q) \, 
e^{-(-ix)^2/4t} e^{-q^2/4t} E_\mu \left( \dfrac{-ix}{(2t)^{1/2}}, \dfrac{q}{(2t)^{1/2}}  \right)
 \psi (q) \nonumber
\\
\label{dt}
= e^{x^2/4t} \mathcal{F}_{\mu, 2t} \left( e^{-(\cdot)^2/4t} \psi(\cdot) \right)(x),
\end{gather}
where $\mathcal{F}_{\mu, 2t}$ is the Dunkl transform.
(See \cite{dJ, DU, RO1, RO2} for information on this transform and \cite{SBS2}
for our notation and conventions.
For this argument, we only need to know that
$$
     \mathcal{F}_{\mu, t} : L^2 (\mathbb{R}^N, \omega_{\mu, t}) \to
 L^2 (\mathbb{R}^N, \omega_{\mu, t})
$$
is injective.)
At this point, let us note that
$\psi \in \mathrm{Dom}  (R^*) \subset L^2(\omega_{\mu,t})$
implies that $ e^{-(\cdot)^2/4t} \psi (\cdot) \in L^2(\omega_{\mu,2t})$ so that
equation (\ref{dt}) makes sense.

We now assume that $\psi \in \mathrm{Ker} \, R^* \subset  \mathrm{Dom} (R^*)$.
So, $R_{\mu, t}^* \psi (- i x) = 0$ for all $x \in \mathbb{R}^N$.
Using that $\mathcal{F}_{\mu, 2t}$ is injective
on $ L^2(\omega_{\mu,2t})$, it follows from (\ref{dt}) that
$\psi = 0$ almost everywhere with respect to the measure $\mathrm{d}\omega_{\mu, 2t}$.
Hence $\psi = 0$ almost everywhere with respect to $\mathrm{d}\omega_{\mu, t}$.
This shows that $R_{\mu, t}^*$ is injective.

To prove that the ranges are dense we will again use Proposition \ref{fa_prop}.
Since $R_{\mu, t}$ is closed we have that
$\overline{ \mathrm{Ran} \, R_{\mu, t}^*  } = ( \mathrm{Ker} \, R_{\mu, t}  )^\perp
= 0^\perp = \mathcal{C}_{\mu, t}$ and that
$( \mathrm{Ran} \, R_{\mu, t}  )^\perp = \mathrm{Ker} \, R_{\mu, t}^* = 0$.
The last equality then implies that
$ \overline{ \mathrm{Ran} \, R_{\mu, t} } =
( \mathrm{Ran} \, R_{\mu, t}  )^{\perp\perp} = 0^\perp =
L^2 (\mathbb{R}^N, \omega_{\mu, t})$.
(We use the symbol $0$ here to designate ambiguously the zero subspace of
the appropriate Hilbert space.)
\hskip 0.5cm
$\blacksquare$

We have shown that the range of the restriction operator
$R_{\mu, t}$ is dense only for the sake of completeness.
This will not be used later on.

We continue with our main result.

\begin{theorem} (Restriction Principle: Version $C$)\\
(i) Suppose that the multiplicity function satisfies $\mu \ge 0$.
 The restriction principle says that the partial isometry $U_{\mu,t}$ produced
by writing the adjoint of the restriction operator, namely $R_{\mu,t}^* $, in its polar
decomposition, that is,
$$
     R_{\mu,t}^* = U_{\mu,t} | R_{\mu,t}^*  |,
$$
actually gives a unitary isomorphism
$U_{\mu,t}:  L^2 ( \mathbb{R}^N, \omega_{\mu,t} ) \to  \mathcal{C}_{\mu,t} $.
\\
(ii) Moreover, we have that
$$
      U_{\mu,t} = C_{\mu,t},
$$
where $C_{\mu, t}$ is defined by equation (\ref{def_C}).

So it follows that
$C_{\mu,t} : L^2 ( \mathbb{R}^N, \omega_{\mu,t} ) \to \mathcal{C}_{\mu,t} $, the
$C$-version of the Segal-Bargmann transform associated with a finite Coxeter group
and the value $t>0$ of Planck's constant, is a unitary isomorphism.
\end{theorem}
\textbf{Remark:} Instead of using the definition (\ref{def_C}) from \cite{SBS2},
we can use the first part of this theorem to define $C_{\mu, t}  := U_{\mu, t}$.
It is in this sense that the restriction principle can be said to define
the $C$-version of the Segal-Bargmann transform.

\vskip 0.2cm \noindent
\textbf{Proof:}
We begin by finding another formula for $R_{\mu, t}^* = R^*$.
So we take $\psi \in \mathrm{Dom} (R^*) \subset L^2(\omega_{\mu,t})$
and $z \in \mathbb{C}^N$.
Continuing the calculation given above in equation (\ref{Rstar}), we obtain
\begin{eqnarray*}
R^* \psi (z)
&=& \int_{ \mathbb{R}^N } \mathrm{d} \omega_{\mu, 2t}(q) \, \rho_{\mu, 2t } (z, q ) \psi (q)
\\
&=& (e^{2t \Delta_\mu/2} \psi) (z) = (e^{t \Delta_\mu} \psi) (z).
\end{eqnarray*}
(Parenthetically, we warn the reader that this equation does \textit{not} say
that $R_{\mu, t}^*$ is equal to $e^{t \Delta_\mu}$.
This quite simply can not be true, since the codomains of these two operators are
not the same space.
The correct statement is that $R_{\mu, t}^*$ is equal to $e^{t \Delta_\mu}$ followed by
analytic continuation to $\mathbb{C}^N $.
Also, it is clear that $R_{\mu, t}^* = C_{\mu, 2t} $, since the domains of $R_{\mu, t}^*$
and $ C_{\mu, 2t} $ are equal as \textit{sets}.)

To get the polar decomposition of $R^*$ we have to analyze the operator
$R^{**} R^*$. But $R^{**} = \overline{R} = R$, since $R$ is closed.
So we consider $R R^*$ from now on.
By using the definition of $R$
we immediately get for $x \in \mathbb{R}^N$ and $\psi \in \mathrm{Dom} \, (R R^*)$ that
$$
    ( R R^* \psi) (x) = (e^{t \Delta_\mu} \psi) (x)
$$
and so
$$
   R R^* = e^{t \Delta_\mu}
$$
on $\mathrm{Dom} \, (R R^*)$ which is dense in $L^2(\mathbb{R}^N, \omega_{\mu, t})$
by a theorem of von~Neumann.
(See \cite{KA}, Chap.~5, \S 3, Thm.~3.24, p.~275.)
But $R R^* $ is closed (being self-adjoint by standard functional analysis)
and bounded (being a restriction
of the bounded operator $e^{t \Delta_\mu}$) and so is a globally defined, bounded
operator by Proposition \ref{fa_prop}.
Moreover, $R R^* = e^{t \Delta_\mu}$ on $L^2(\omega_{\mu, t})$.
So, $| R^* | = ( R R^* )^{1/2}$ is a globally defined, bounded operator with
$| R^* | = ( e^{t \Delta_{\mu} } )^{1/2} =
e^{t \Delta_\mu/2}$ on $L^2 (\mathbb{R}^N, \omega_{\mu,t} ) $, since the operator
$e^{t \Delta_\mu/2} \ge 0$, is globally defined, bounded and its square is $e^{t \Delta_\mu}$.

Next the polar decomposition theorem tells us that
\begin{equation}
\label{Rstar_decomp}
       R^* = U_{\mu, t} \, | R^* |
\end{equation}
on $\mathrm{Dom} (R^*) = \mathrm{Dom} (| R^* |)$,
where $U_{\mu, t}$ is partial isometry from $L^2 ( \omega_{\mu,t} ) $ to
$\mathcal{C}_{\mu, t}$.
But $\mathrm{Dom} (| R^* |) = L^2 ( \omega_{\mu,t} )$ and so
$R^*$ is globally defined and equal by (\ref{Rstar_decomp})
to the composition of two bounded operators on $L^2 (\mathbb{R}^N, \omega_{\mu,t} )$.
Therefore $R^*$ is also bounded.
Since $R$ is closed, we have $R = \overline{R} = (R^*)^*$.
This displays $R$ as the adjoint of the globally defined, bounded operator $R^*$.
We then conclude that $R$ is a globally defined, bounded operator as well.

Now by a ``one-page'' argument we have shown that $\mathrm{Ker} \, R^* = 0 $,
and so $U_{\mu, t} $ is one-to-one.
And by a ``one-line'' proof we have seen that $\mathrm{Ran} \, R^*$ is dense,
and so $U_{\mu, t} $ is onto.
The two preceding assertions about $U_{\mu, t}$ follow
from the Polar Decomposition Theorem \ref{polar_decomp}.
We conclude that $U_{\mu, t} $ is a unitary isomorphism.

We now write equation (\ref{Rstar_decomp}) equivalently as
$$
(e^{t \Delta_\mu} \psi) (z) = ( U_{\mu, t} e^{t \Delta_\mu/2} \psi ) (z)
$$
for all $\psi \in L^2(\omega_{\mu,t})$ and all $z \in \mathbb{C}^N$.
Now we apply $R_{\mu, t}$ to both sides, recalling that there is an implicit analytic
continuation on the left side which cancels with  $R_{\mu, t}$, to get
$$
(e^{t \Delta_\mu} \psi) (x) = ( R_{\mu, t}  U_{\mu, t} e^{t \Delta_\mu/2} \psi ) (x)
$$
for all $x \in \mathbb{R}^N$ and all $\psi \in L^2(\omega_{\mu,t})$.
So, we have the operator equation
$$
e^{t \Delta_\mu}  =  R_{\mu, t}  U_{\mu, t} e^{t \Delta_\mu/2}
$$
where each side is a bounded operator from $L^2(\omega_{\mu,t})$ to itself.
Also all of the operators in this equation are bounded.
This then implies that
$$
e^{t \Delta_\mu} e^{-t \Delta_\mu/2} =  R_{\mu, t}  U_{\mu, t}
$$
on $\mathrm{Ran} \, ( e^{t \Delta_\mu/2} ) \subset L^2 ( \omega_{\mu,t} ) $.
Of course, $ e^{-t \Delta_\mu/2}$ is not a bounded operator.
However, its domain $\mathrm{Ran} \, e^{t \Delta_\mu/2}$ is dense
in $L^2 ( \omega_{\mu,t} ) $.
(Proof: Using the Dunkl transform $\mathcal{F}_{\mu, t}$ (see \cite{dJ, DU, RO1, RO2})
one shows that the bounded operator $e^{t \Delta_\mu/2}$
is unitarily equivalent to multiplication by $e^{ - t k^2 /2}$ acting
on $ L^2 ( \mathbb{R}^N, \omega_{\mu, t})$, where $k$ is the variable in $\mathbb{R}^N$.
But the range of multiplication by $e^{ - t k^2 /2}$ clearly contains
$C^\infty_0 (\mathbb{R}^N)$ and so is dense by a standard argument in analysis.)
Moreover, we also have
$$
e^{t \Delta_\mu} e^{-t \Delta_\mu/2} =   e^{t \Delta_\mu/2}
$$
on $\mathrm{Ran} \, ( e^{t \Delta_\mu/2} )$ as one sees by applying both sides
to an arbitrary element
$\phi = e^{t \Delta_\mu/2} \psi \in \mathrm{Ran} \, ( e^{t \Delta_\mu/2} )$,
where $\psi \in L^2(\omega_{\mu, t})$,
and by using the semi-group property.
This in turn gives us
$$
e^{t \Delta_\mu/2} =  R_{\mu, t}  U_{\mu, t}
$$
on the dense domain $\mathrm{Ran} \, ( e^{t \Delta_\mu/2} )$.
Since both $e^{t \Delta_\mu/2}$ and $R_{\mu, t}  U_{\mu, t}$
are globally defined, bounded operators that are equal on a dense domain,
it follows that
$$
e^{t \Delta_\mu/2} =  R_{\mu, t}  U_{\mu, t}
$$
on $L^2 ( \omega_{\mu,t} ) $.
So for all
$x \in \mathbb{R}^N$ and all $\psi \in L^2(\omega_{\mu,t})$, we obtain
$$
( e^{t \Delta_\mu/2} \psi ) (x) =  (R_{\mu, t} U_{\mu, t} \psi ) (x).
$$
Next, we write out the left side as follows:
$$
(e^{t \Delta_\mu/2} \psi) (x) =
\int_{ \mathbb{R}^N } \mathrm{d} \omega_{\mu, t} (q) \rho_{\mu, t} (x,q) \psi (q)
=  (R_{\mu, t} C_{\mu, t} \psi) (x).
$$
So for all
$x \in \mathbb{R}^N$ and all $\psi \in L^2(\omega_{\mu,t})$, we find that
$$
(R_{\mu, t}  U_{\mu, t} \psi) (x) =  (R_{\mu, t} C_{\mu, t} \psi) (x)
$$
and so $R_{\mu, t}  U_{\mu, t} =  R_{\mu, t} C_{\mu, t}$.
Using that $R_{\mu, t}$ is injective (that is, uniqueness of analytic
continuation) we finally arrive at the desired identity,
$
  U_{\mu, t} =  C_{\mu, t},
$
and therefore $C_{\mu, t}$ is a unitary isomorphism as we wanted to prove.
\hskip 1cm $\blacksquare$

During the proof of the previous theorem we proved the statement made earlier
that $R_{\mu, t}$ is bounded.
We now state this result separately and amplify on it.
\begin{theorem}
The operator $R \equiv R_{\mu, t}$ is bounded and has operator norm $|| R || = 1$.
Also the operator $R^*$ is bounded with operator norm $|| R^* || = 1$.
\end{theorem}

\noindent
\textbf{Proof:}
In this proof we denote all operator norms by $|| \cdot ||$.
We already have shown that $ | R^* |^2 = R R^* $ is a self-adjoint,
bounded operator acting on  $L^2 ( \mathbb{R}^N, \omega_{\mu,t} )$ and that
$R$ and $R^*$ are globally defined, bounded operators.
We take $ \phi \in L^2 ( \mathbb{R}^N, \omega_{\mu,t} )$ in the following, getting
\begin{eqnarray*}
 || \,\, |R^*|^2 \, || &=& \sup_{||\phi||=1}
\langle \phi,|R^*|^2 \phi \rangle_{L^2 ( \omega_{\mu,t} )}
 =
\sup_{||\phi||=1} \langle \phi,  R R^* \phi \rangle_{L^2 ( \omega_{\mu,t} )}
\\
&=& \sup_{||\phi||=1} \langle R^* \phi,  R^* \phi \rangle_{  \mathcal{B}_{\mu, t} }
 = \sup_{ ||\phi||=1} ||  R^* \phi ||^2_{  \mathcal{B}_{\mu, t} }
\\
&=& || R^* ||^2 = || R ||^2.
\end{eqnarray*}
We also compute directly
$$
    || \,\, |R^*|^2 \, || = || ( e^{t \Delta_\mu/2 })^2 || = || e^{t \Delta_\mu} || = 1,
$$
since $\mathrm{Spec} ( \Delta_{\mu} ) = ( -\infty, 0]$ and $t > 0$.
The result now follows.
\hskip 0.5cm $\blacksquare$

\section{Versions $A$, $B$ and $D$}

Now we will apply the method indicated after the statement of the
Polar Decomposition Theorem \ref{polar_decomp} in order to show that the $A$-version of
the Segal-Bargmann transform can be obtained by a polar decomposition
which is related to the polar decomposition (namely, the restriction principle)
used to obtain the $C$-version.
So, we are looking for two unitary isomorphisms, $F_{1}$ and $F_{2}$,
making the following diagram commute:
\begin{equation}
\label{comm_diagram}
\begin{array}{ccc}
L^2(\mathbb{R}^N, \omega_{\mu,t}) & \stackrel{{F_1}}{\longrightarrow} &
 L^2( \mathbb{R}^N,  \omega_{\mu,t}) \\
\Big\downarrow\vcenter{ \llap{ $C_{\mu,t}~~$ }  }  & & 
\Big\downarrow\vcenter{ \rlap{ $A_{\mu,t}$ }  } \\
\mathcal{C}_{\mu,t} &
\stackrel{{F_2}}{\longrightarrow} & \mathcal{B}_{\mu,t}
\end{array}
\end{equation}
Then we can use these two unitaries to change
the polar decomposition which gave us $C_{\mu,t}$
into a polar decomposition giving $A_{\mu,t}$.
Of course, the very existence of such a pair, $F_{1}$ and $F_{2}$,
already would prove that $A_{\mu, t}$ is a unitary isomorphism.

We use a known relation between the $A$ and $C$-versions in order to start.
The rest of the construction then follows in a systematic, algorithmic manner.
The relation between these two versions that we use starts from this identity
for the integral kernels:
$$
C_{\mu, t} ( 2z, q ) = A_{\mu, 2t} ( 2z, 0) A_{\mu, t/2} ( z, q) =
                       e^{-z^2/t}A_{\mu, t/2} ( z, q)
$$
for all $z \in \mathbb{C}^N$ and all $q \in \mathbb{R}^N$.
(See \cite{SBS2}, Theorem 3.5.)
Now we translate this relation into a relation between the integral
transforms themselves.
From the defining equation (\ref{def-kernel-a}) we have the scaling relation
$
A_{\mu, \lambda^2 t} (\lambda z, \lambda q ) = A_{\mu, t} ( z, q )
$
for $\lambda > 0$.
By taking $\lambda = 2^{1/2}$ and replacing $t$ with $t/2$ in this, we have
$$
C_{\mu, t}( 2z, q )=e^{-z^2/t}A_{\mu, t/2}( z, q)=e^{-z^2/t}A_{\mu, t}( 2^{1/2}z, 2^{1/2}q).
$$
Next we replace $z$ with $ 2^{-1/2} z$ to obtain
\begin{equation}
\label{CA_rel}
C_{\mu, t}( 2^{1/2}z, q )= e^{-z^2/2t} A_{\mu, t}( z, 2^{1/2}q).
\end{equation}
To understand the integral kernel $A_{\mu, t}( z, 2^{1/2}q)$ we take
$\psi \in L^2(\mathbb{R}^N, \omega_{\mu,t} )$ and evaluate as follows:
\begin{gather*}
\int_{\mathbb{R}^N} \mathrm{d} \omega_{\mu,t} (q)  A_{\mu, t}( z, 2^{1/2}q) \psi (q) =
\int_{\mathbb{R}^N} \mathrm{d} \omega_{\mu,t} ( 2^{-1/2} \tilde{q})
A_{\mu, t}( z, \tilde{q}) \psi (2^{-1/2}\tilde{q})
\\
 =
2^{-(\gamma_{\mu} +N/2)}  \int_{\mathbb{R}^N} \mathrm{d} \omega_{\mu,t} ( \tilde{q})
A_{\mu, t}( z, \tilde{q}) D_{2^{-1/2}} \psi (\tilde{q})
= 2^{-(\gamma_{\mu} +N/2)} ( A_{\mu, t}  D_{2^{-1/2}} \psi ) (z),
\end{gather*}
where we used $\tilde{q} = 2^{1/2} q$, the scaling property
$\omega_{\mu, t}(\lambda q) = |\lambda|^{2\gamma_{\mu}} \omega_{\mu, t}(q) $,
$\mathrm{d}^N q = 2^{-N/2} \mathrm{d}^N \tilde{q}$ and
the definition of the dilation operator $D_{2^{-1/2}}$.
Next, by multiplying both sides of (\ref{CA_rel}) by
$\psi \in L^2(\mathbb{R}^N, \omega_{\mu,t} )$  and then integrating with respect to
$\mathrm{d} \omega_{\mu,t} (q)$, we get
$$
 C_{\mu, t} \psi (2^{1/2}z) = 2^{-(\gamma_{\mu} +N/2)}
e^{-z^2/2t} ( A_{\mu, t}  D_{2^{-1/2}} \psi ) (z)
$$
A crucial point here is that the factor $e^{-z^2/2t}$ does not depend
on the variable of integration and so factors out in front of of the integral.
Equivalently,
$$
    D_{2^{1/2}} C_{\mu, t} \psi (z) =
 2^{-(\gamma_{\mu} +N/2)}
e^{-z^2/2t} ( A_{\mu, t}  D_{2^{-1/2}} \psi ) (z),
$$
which itself is equivalent to the operator equation
$$
    D_{2^{1/2}} C_{\mu, t}  =  2^{-(\gamma_{\mu} +N/2)}
        e^{-(\cdot)^2/2t} A_{\mu, t}  D_{2^{-1/2}},
$$
where $ e^{-(\cdot)^2/2t}$ denotes the operator of multiplication
by the function $ e^{-z^2/2t}$.
Now we solve the last equation for $A_{\mu, t}$ getting
\begin{equation}
\label{sandwich}
A_{\mu, t} =  2^{\gamma_{\mu} +N/2 }
         e^{(\cdot)^2/2t}  D_{2^{1/2}} C_{\mu, t} D_{2^{1/2}}.
\end{equation}

Next, we want the operator $C_{\mu, t}$ to be sandwiched between two unitary
operators, and so it is not initially clear how to divide up the factors
of $2$ in equation (\ref{sandwich}) to get multiples of $ D_{2^{1/2}} $ and of
$ e^{ (\cdot)^2/2t}  D_{2^{1/2}}$ that are unitaries.
But by Lemma \ref{dilation_lemma} we know that
$$
2^{\gamma_{\mu}/2 + N/4} D_{2^{1/2}} : L^2 (\mathbb{R}^N, \omega_{\mu, t}) \to
L^2(\mathbb{R}^N, \omega_{\mu, t})
$$
is a unitary isomorphism.
The desired domain and the desired codomain of this unitary operator are
determined by diagram (\ref{comm_diagram}).
So it remains to show what is happening with the operator
$F_2 := 2^{\gamma_{\mu}/2 + N/4}  e^{(\cdot)^2/2t}  D_{2^{1/2}} $.
According to the diagram (\ref{comm_diagram}) this should be the unitary
isomorphism $F_2 : \mathcal{C}_{\mu, t} \to \mathcal{B}_{\mu, t}$
indicated there.

Therefore we would like to take $f \in \mathcal{C}_{\mu, t} $ and calculate
the norms $|| f ||_{ \mathcal{C}_{\mu, t} }$ and
 $|| F_2 f ||_{ \mathcal{B}_{\mu, t} }$ and then show they are equal.
But we do not have closed formulas for these norms for general elements
in these reproducing kernel Hilbert spaces.
However, it suffices to consider the case when $f = L_z \in \mathcal{C}_{\mu, t}$,
where $z \in \mathbb{C}^N$ is arbitrary.
See (\ref{defineL}).
In spite of the quantity of details, this does work out in an algorithmic manner.

Nevertheless, purely for the sake of simplicity, we prefer to give a shorter proof
by relating $F_2$ with known entities.
We note first that $G: \mathcal{C}_{\mu, t} \to \mathcal{B}_{\mu, t/2}$
is a unitary isomorphism.
(See (\ref{define_G}) and the subsequent discussion.)
And second from a result in \cite{SBS2} we have that
$D_{2^{-1/2}} :\mathcal{B}_{\mu, t/2} \to \mathcal{B}_{\mu, t}$
is also a unitary isomorphism.
So the composition
$
   D_{2^{-1/2}} \, \, G : \mathcal{C}_{\mu, t} \to \mathcal{B}_{\mu, t}
$
is again a unitary isomorphism.
For any $f \in \mathcal{C}_{\mu, t}$ we use (\ref{define_G}) to calculate this composition,
giving for all $w \in \mathbb{C}^N$ that
\begin{gather*}
   ( D_{2^{-1/2}} \, \, G f )(w) = Gf ( 2^{-1/2} w ) =
2^{\gamma_{\mu}/2  + N/4} f ( 2 \cdot 2^{-1/2} w) e^{ ( 2^{-1/2} w )^2 / t}
\\
= 2^{\gamma_{\mu}/2  + N/4} f ( 2^{1/2} w) e^{w^2 / 2t}
= 2^{\gamma_{\mu}/2  + N/4} e^{w^2 / 2t} (D_{2^{1/2}} f)(w),
\end{gather*}
which in turn implies the operator equation
$$
  D_{2^{-1/2}} \, \, G =  2^{\gamma_{\mu}/2  + N/4} e^{(\cdot)^2 / 2t} D_{2^{1/2}} = F_2.
$$
It follows that
$
F_2 : \mathcal{C}_{\mu, t}
\to \mathcal{B}_{\mu, t}
$
is a unitary isomorphism.

We are now ready to apply the method discussed in the remarks just after the
Polar Decomposition Theorem \ref{polar_decomp}.
Using the notation established there, we let
$$
      F_1 : L^2 (\mathbb{R}^N, \omega_{\mu, t}) \equiv \mathcal{H}_1 \to
L^2 (\mathbb{R}^N, \omega_{\mu, t}) \equiv \mathcal{K}_1
$$
be defined as
$$
F_1 := (2^{\gamma_{\mu}/2 + N/4} D_{2^{1/2}})^*
:= (2^{\gamma_{\mu}/2 + N/4} D_{2^{1/2}})^{-1}
= 2^{-(\gamma_{\mu}/2 + N/4)} D_{2^{-1/2}}.
$$
Also, we already defined $F_2= 2^{\gamma_{\mu}/2  + N/4} e^{(\cdot)^2 / 2t} D_{2^{1/2}} $.
So we have shown above that $F_1$ and $F_2$ are unitary isomorphisms
and that diagram (\ref{comm_diagram}) commutes.

Of course, we have from equation (\ref{sandwich}) and the subsequent results
that $A_{\mu, t} = F_2 C_{\mu, t} F_1^*$ is a unitary isomorphism, since it is
the composition of three unitary isomorphisms.
We now want to see how $A_{\mu, t}$ arises explicitly from the corresponding
polar decomposition (which, according to our definition, will turn out not
to be a restriction principle) and how this polar decomposition relates
to the unmotivated definition of a ``restriction'' operator in \cite{SBSBO}.
So, continuing with the notation established earlier we have that $A_{\mu, t}$
arises in the polar decomposition $B = V |B|$, that is $V = A_{\mu, t}$,
where $B = F_2 R_{\mu, t}^* F_1^*$.
(Recall that we have shown that $R_{\mu, t}$ and $R_{\mu, t}^*$ are globally defined,
bounded operators.)
It follows that $B^* = F_1 R_{\mu, t} F_2^*$ and therefore $A_{\mu, t}$ arises
from the restriction principle according to our definition exactly when
$F_1 R_{\mu, t} F_2^* $ is the restriction operator
$\mathcal{B}_{\mu, t} \to L^2(\mathbb{R}^N, \omega_{\mu, t})$, namely,
$f \mapsto f \upharpoonright_{\mathbb{R}^N}$.
We know that
$F_2 =  2^{\gamma_{\mu}/2  + N/4} e^{(\cdot)^2 / 2t} D_{2^{1/2}} = D_{2^{-1/2} } \, \, G $
and so $F_2^* = F_2^{-1} = G^{-1} \,  D_{2^{1/2} }$.
But from (\ref{define_G}) we immediately have
$$
     G^{-1} g (w) = 2^{ -(\gamma_{\mu}/2 +N/4) } e^{-w^2/ 4t} g (w/2) .
$$
So for $f \in \mathcal{B}_{\mu, t}$ we have for $ w \in \mathbb{C}^N$ that
\begin{gather*}
F_2^* f (w) =  ( G^{-1} \, D_{2^{1/2} } f ) (w) =
2^{ -(\gamma_{\mu}/2 +N/4) }  e^{-w^2/ 4t} (D_{2^{1/2}} f) (w/2)
\\
= 2^{ -(\gamma_{\mu} +N/4) } e^{-w^2/ 4t} f ( 2^{-1/2} w) .
\end{gather*}
Then since $R_{\mu, t}$ is simply restriction, we obtain for all $x \in \mathbb{R}^N$ that
\begin{gather*}
     (R_{\mu, t} F_2^* f ) (x)
= 2^{ -(\gamma_{\mu}/2 +N/4) } e^{-x^2/ 4t} f ( 2^{-1/2} x) .
\end{gather*}
Finally, applying $F_1 = 2^{-(\gamma_{\mu}/2 + N/4)} D_{2^{-1/2}}$
yields for all $f \in \mathcal{B}_{\mu, t}$ and $x \in \mathbb{R}^N$
\begin{equation}
\label{not_a_restrn}
    (F_1 R_{\mu, t} F_2^* f ) (x) = 2^{ -(\gamma_{\mu} +N/2) } e^{-x^2/ 8t} f(x/2),
\end{equation}
which is \textit{not} the restriction operator.
Consequently, this polar decomposition is not a restriction principle.
However, notice that the operator $F_1 R_{\mu, t} F_2^*$ is globally defined
and bounded, since $R_{\mu, t}$ is globally defined and bounded.
This fact is not so obvious by merely inspecting the right side of (\ref{not_a_restrn}).

The operator in (\ref{not_a_restrn})
does not compare very well at first sight
with the ``restriction operator'' defined in \cite{SBSBO}, p.~298.
But this discrepancy is easily understood.
In Corollary 3.1 in \cite{SBS2} we give the unitary equivalence between
$A_{\mu, 1}$ and the ``generalized Segal-Bargmann transform'' $BSO$ defined in \cite{SBSBO}.
(N.B. Only the case $t=1$ is considered in \cite{SBSBO}.)
Using this we can conjugate the polar decomposition used above in order to obtain
$A_{\mu, 1}$ to get an operator, say $S$, whose polar decomposition gives us $BSO$.
We note that $S$ is  globally defined and bounded,
since it is unitarily equivalent to $R_{\mu, t}^*$.
The adjoint of $S$ (which should be the restriction operator)
for all $f \in \mathcal{B}_{\mu, t}$ and $x \in \mathbb{R}^N$ turns out to be
$$
S^* f (x) = c_{\mu}^{-1/2} e^{-x^2 / 2} f(x),
$$
which is not a
restriction operator according to our definition.
Except for the positive multiplicative constant $ c_{\mu}^{-1/2}$,
this agrees with the ``restriction operator'' given in  \cite{SBSBO}.
But for any closed, densely defined operator $T$ and any $\lambda > 0$,
the polar decompositions of $T$ and $\lambda T$ give the same partial isometry.
And this explains how the unmotivated ``restriction operator''
used in \cite{SBSBO} arises in a natural manner in our presentation.

We wish to note that formula (\ref{not_a_restrn}) was forced on us by our
method, once we had established that the unitary operators $F_1$ and $F_2$
change the transform $C_{\mu, t}$ into $A_{\mu, t}$.
(Cp. diagram (\ref{comm_diagram}).)
And these two unitaries arose in a natural, motivated way directly from an
identity that relates the kernel functions of these transforms.
So the $A$-version arises by applying polar decomposition
to a particular operator.
When one thinks of it this way, this is a rather unimpressive result.
Actually, \textit{every} unitary operator between two Hilbert spaces
can be realized via a polar decomposition.
And \textit{any} closed, densely defined operator which satisfies two
additional hypotheses (injectivity and dense range) gives us a unitary operator
in its polar decomposition.

Moreover, we could have used another pair of unitary isomorphisms, say $G_1$ and
$G_2$ in place of $F_1$ and $F_2$,
to change  $C_{\mu, t}$ into  $ Z := G_2 C_{\mu, t} G_1^*$,
using a diagram analogous to (\ref{comm_diagram}).
Then  $Z$ arises from the polar decomposition that comes from the restriction
principle used to produce $C_{\mu, t}$.
However, this polar decomposition in general will not be a restriction principle.
(For example, the codomains of $G_1$ and $G_2$ need not even be function spaces.)
Actually,
\textit{any} unitary isomorphism $Z$ between separable, complex Hilbert spaces
of infinite dimension can arise this way by an appropriate, but far from unique, choice
of the two unitaries $G_1$ and $G_2$.
So in general it would be misleading to dub $Z$ with a name that indicates that
it forms a part of Segal-Bargmann analysis.

However, the transform $A_{\mu, t}$ does arise naturally and uniquely from
the heat kernel method as a part of Segal-Bargmann analysis.
(See \cite{SBS2}.)
So it is reasonable to ask (and answer, as we have done in this section)
how the restriction principle for $C_{\mu, t}$
gives us a polar decomposition of $A_{\mu, t}$.
On the other hand,
we have not been able to find in \cite{SBSBO} a satisfactory, explicit justification
for considering the transform defined there as a part of Segal-Bargmann analysis.
For example, Remark 4.3 (\cite{SBSBO}, p.~301) only indicates what
happens when $\mu \equiv 0$ (in our notation).
In our opinion this is very far from justifying the terminology
``Segal-Bargmann'' for the case of general $\mu$.

One point of this section is to show where the unmotivated exponential
factor comes from in the definition of the ``restriction operator'' in \cite{SBSBO}.
It is truly a \textit{deus ex machina} in \cite{SBSBO}.
Here it flows out naturally from an analysis based on the $C$-version.
The second point of this section is to provide contrast with the
method used to define the $C$-version in the last section.
While that was also a polar decomposition, it was a particular, uniquely
defined special case, namely the restriction principle.
The worst that could happen with an analysis based on the restriction principle
is that the technical details do not work out and therefore no
unitary isomorphism at all is produced.
In short, the result of the method is unique, but may not exist.

As for the remaining two versions of the Segal-Bargmann, the Version $B$
(resp. $D$) is defined by a unitary transformation (a change of measure)
on the domain space starting with the Version $A$ (resp. $C$).
(See \cite{SBS2} for details about Version~$B$ and \cite{SBS3} for Version~$D$.)
So the restriction principle for the $C$-version implies that these remaining two versions
can also be obtained from the polar decomposition of an explicitly defined operator.
The details are left to the interested reader.
We do wish to comment that these polar decompositions are not restriction principles.
The brevity of our discussion in this paragraph is not meant to indicate
that these versions are less important than the $A$-version.
On the contrary, we think that the three versions $A$, $B$ and $D$ have the
same relative relation to the truly important and logically central $C$-version.

\section{Concluding Remarks}

Our confusion over the role in \cite{SBSBO}
of their ``restriction principle'' in Segal-Bargmann analysis motivated our
study of this topic.
The upshot is our discovery of the central role of the restriction principle
in the $C$-version of Segal-Bargmann analysis associated to a finite Coxeter group.
We wish to underscore that only the $A$-version of Segal-Bargmann analysis
is considered in \cite{SBSBO}.
This can be clearly seen in the reproducing kernel for the space of
holomorphic functions in \cite{SBSBO}, which is therefore the $A$-version space.
Also the ``generalized Segal-Bargmann transform'' in \cite{SBSBO} has an integral kernel
which is not the analytically continued heat kernel (as in the
$C$-version), but rather something that corresponds to our uniquely defined
$A$-version (modulo normalization and dilation).
There is no mention in \cite{SBSBO} of the $C$-version nor even of the existence
of other versions of Segal-Bargmann analysis.

In summary, we think that this paper shows
that the restriction principle and the $C$-version
(and not any other version) of Segal-Bargmann analysis
are naturally and closely related with each other.
So this is a new way for understanding how the $C$-version in general
is the most fundamental version of Segal-Bargmann analysis.

As for future endeavors, we note that
we have studied only the case $\mu \ge 0 $ and so it might be
interesting to understand what happens when we drop or weaken that condition.

\section{Acknowledgments} My thanks go to Greg Stavroudis for
a most hospitable venue for connecting to the Internet as well as for his excellent coffee,
both being critical elements for doing the research for this paper.

\end{document}